\documentclass[%
 reprint,
 amsmath,amssymb,
 aps,
 prl,
]{revtex4-2}

\usepackage{graphicx}
\usepackage{dcolumn}
\usepackage{bm}
\usepackage{amsmath}
\usepackage{physics}
\usepackage{mathtools}
\usepackage{float}
\usepackage[subrefformat=parens]{subcaption}
\usepackage{listings}
\usepackage{here}
\usepackage{siunitx}
\usepackage{multirow}
\usepackage{ulem}
\usepackage{diagbox}
\usepackage{lineno}

\DeclareRobustCommand{\erase}{\bgroup\markoverwith{\textcolor{red}{\rule[.5ex]{2pt}{0.4pt}}}\ULon}

\usepackage{xcolor}

\begin{document}

\preprint{APS/123-QED}
\title{Improving the Rate-Loss Scaling in Polarization Entanglement Distribution using Single-Click Entanglement Swapping}

\author{Hikaru Shimizu$^{1}$}
 \email{shimihika2357@keio.jp}
 
\author{Joe Yoshimoto$^{2,3}$}

\author{Daiki Ichii$^{1}$}

\author{Junko Ishi-Hayase$^{2,3}$}
 
\author{Rikizo Ikuta$^{4,5}$}
 \email{ikuta.rikizo.es@osaka-u.ac.jp}

\author{Masahiro Takeoka$^{1,6}$}
 \email{takeoka@elec.keio.ac.jp}

\affiliation{%
$^{1}$\mbox{Department of Electronics and Electrical Engineering, Keio University, 
Kohoku-ku, Yokohama 223-8522, Japan}
} 
\affiliation{%
$^{2}$\mbox{School of Fundamental Science and Technology, Keio University, 
Kohoku-ku, Yokohama 223-8522, Japan}
} 
\affiliation{%
$^{3}$Center for Spintronics Research Network, Keio University, 
Kohoku-ku, Yokohama 223-8522, Japan
}
\affiliation{%
$^{4}$Graduate School of Engineering Science, Osaka University, Toyonaka, Osaka 560-8531, Japan
}
\affiliation{%
$^{5}$Center for Quantum Information and Quantum Biology, Osaka University, Osaka 560-0043, Japan
}
\affiliation{%
$^{6}$National Institute of Information and Communications Technology~(NICT), Koganei, Tokyo 184-8795, Japan
}
\date{\today}

\begin{abstract}
Polarization entanglement is widely used in optical quantum information processing due to its compatibility with standard optical components.
On the other hand, it is known that polarization entanglement is susceptible to the loss, more precisely, its transmission rate in a lossy channel is limited by the scaling of $O(\eta_{\rm C})$, where $\eta_{\rm C}$ is the transmittance of the channel. 
Here, we experimentally demonstrate that this rate-loss scaling limit can be overcome by a relatively simple protocol. 
This is possible by integrating the idea of the polarizaion-photon-number hybrid entanglement and the single-click entanglement swapping. 
We demonstrate square root improvement of the rate-loss scaling from the conventional approaches and achieve the fidelity of 0.843 for the distributed polarization entangled photon pairs. 
This improvement in the rate-loss scaling is equivalent to that achieved by 1-hop quantum repeater node.
Our result paves a way to build a near-future quantum network and its applications. 
\end{abstract}
\maketitle


{\it Introduction.$-$}
Photonic polarization entanglement is a fundamental resource for various quantum information processing.
The most notable feature is its compatibility with ordinary optical components.
Due to its practicality and ease of use, polarization entanglement has been widely employed in various experiments:
from fundamental science, e.g. quantum state tomography~\cite{White99, James01} and Bell tests~\cite{Giustina15, Tsujimoto18, Liu22}, to applications of quantum information such as 
entanglement-based quantum key distribution~\cite{Shi20, Yin20}, quantum sensing~\cite{Kim24, Zhao21, Liu21}, quantum computing~\cite{Gasparoni04, Walther05, Chen07, Tokunaga08, Crespi11}, as well as quantum network\cite{Craddock24}.

One of the practical issues of the polarization entanglement is the resistance to the channel loss.  
For an optical channel with transmittance $\eta_{\rm C}$, the rate of distributing polarization entangled photons scales linearly with $\eta_{\rm C}$.
This is particularly a problem for long distance quantum communication~\cite{Wengerowsky20, Neumann22} since $\eta_{\rm C}$ decreases exponentially with the distance. 
It is also problematic even for short distance, to generate and apply multi-partite polarization entanglement~\cite{McCutcheon16, Proietti21} since all multipartite photons must be successfully detected simultaneously. 
Experimentally, this is a critical problem since the required time to collect data easily goes to impractically long. 
In principle, quantum repeater~\cite{Briegel98, Sangouard11, Azuma23} can overcome this problem and its technology is growing remarkably ~\cite{Hasegawa19, Li19, Krutyanskiy23, Liu24, Stolk24, Hanni25}. 
However, it is still not easy to leverage the rate-loss scaling advantage of quantum repeaters in real experiments. 


In this paper, we experimentally demonstrate that the rate-loss scaling of distributing polarization entangled photon pairs through a lossy optical channel with transmittance $\eta_{\rm C}$ can be better than $O(\eta_{\rm C})$. 
This is achieved by combining the follwoing two key ideas.
First, we employ a single-click entanglement swapping, which utilizes the superposition of the vacuum and single-photon states.
This approach is known to surpass the scaling of direct transmission~\cite{Campbell08, Lucamarini18} and can also be extended to the distribution of multi-partite entanglement~\cite{Roga23, Shimizu25}.
Several experimental demonstrations of this physical encoding have been reported~\cite{Caspar20, Zo24}.

Second, we use the hybrid entanglement between polarization and photon-number qubits.
This state is obtained by generating normal polarization or photon-number qubit entanglement and converting the degree of freedom~\cite{Fiurasek17} of one of the modes.
An experimental study of hybrid entanglement generation has also been reported~\cite{Drahi21}.
The basic idea of combining these techniques is as follows. 
We prepare hybrid entanglement sources at two end-users, Alice and Bob, and then they send the photon-number superposition parts to lossy channels for the entanglement swapping. 
Then, it leverages the improved rate-loss scaling by the single-click enatnglement swapping whereas the resulted state shared by Alice and Bob is a polarization entangled photon pair. 
Our result paves a way of expanding the practical usefulness of polarization entanglement in quantum network and multi-partite quantum information processings with currently feasible technologies.

\begin{figure}[t]
    \centering
    \includegraphics[width=1\linewidth]{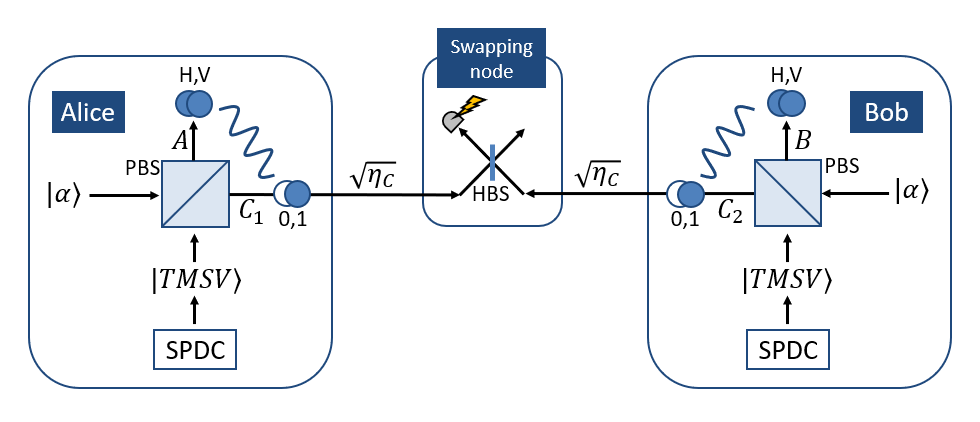}
    \caption{\raggedright Schematic image of the experiment. 
    The Bell state measurement~(BSM) in this protocol is based on the single-photon interference between the number-state photons coming from Alice and Bob. 
    The single-photon detection at the swapping node heralds the successful distribution of the polarization entangled photon pair.}
    \label{fig1}
\end{figure}

{\it Protocol overview.$-$}
The sources of the hybrid entanglement of polarization and photon-number qubits are prepared at both Alice's and Bob's end as shown in Fig.~\ref{fig1}. 
The hybrid entanglement is generated by combining two optical quantum states. One is the two-mode polarization squeezed vacuum~(TMSV) described as, 
\begin{equation}
    \ket{\rm TMSV}\sim \ket{0_H0_V} + \gamma\ket{1_H1_V} + \mathcal{O}(\gamma^2), 
    \label{eq:TMSV}
\end{equation}
for $|\gamma|\ll 1$, where 
$\ket{0_{H(V)}}$ is a vacuum state and $\ket{1_{H(V)}}$ is a single-photon state 
in the H(V)-polarized mode. 
TMSV can be generated by the spontaneous parametric down-conversion~(SPDC) in a Type-$\mathrm{I}\hspace{-1.2pt}\mathrm{I}$ second-order nonlinear optical crystal. 
The other one is vertically polarized weak coherent light described as, 
\begin{equation}
    \ket{\alpha}\sim \ket{0_V} + \alpha\ket{1_V} + \mathcal{O}(\alpha^2), 
    \label{eq:wcp}
\end{equation}
for $|\alpha|\ll 1$. 
They are mixed by using a polarizing beam splitter~(PBS). 
At Alice's side, by post-selecting the events where one or more photons exist in mode A, 
we obtain an unnormalized hybrid state in modes~A and $C_1$ as 
\begin{align}
    \label{initial}
    \ket{\psi}_{AC_1}
    &=
    \alpha\ket{V}_{A}\ket{0}_{C_1}+\gamma\ket{H}_{A}\ket{1}_{C_1}, 
\end{align}
where we omitted terms originating from components of the multiple photons in $\ket{\rm TMSV}$ and $\ket{\alpha}$, which will be discussed later. 
Note that $\ket{H(V)}_{A}$ and $\ket{n_V}_{C_1}$ in Eq.~(\ref{initial}) correspond to $\ket{1_{H(V)}}_{A}$ and $\ket{n}_{C_1}$, respectively, in the previous expressions. 
The same hybrid entanglement is also prepared in Bob's side, denoted as $\ket{\psi}_{BC_2}$.  

The states in modes $C_1$ and $C_2$ are sent to the swapping node through optical channels with transmittance $\sqrt{\eta_{\rm C}}$. 
The swapping is successful if single-photon detection occurs at one of the output of the BS in the swapping node. 
For the successful event, the total state is projected onto 
$\bra{\Psi_{01}^+}_{C1C2}=(\bra{0}_{C1}\bra{1}_{C2}+\bra{1}_{C1}\bra{0}_{C2})/\sqrt{2}$, which acts as the BSM, and the resulting unnormalized state in A and B is given as, 
\begin{align}
\bra{\Psi_{01}^+}_{C1C2}\ket{\psi}_{AC1}\ket{\psi}_{BC2}
= \alpha\gamma\eta_{\rm C}^{1/4}|\Psi_{\rm pol}^+\rangle_{AB}, 
\label{eq:swapping}
\end{align}
where,  
\begin{align}
|\Psi_{\rm pol}^+\rangle_{AB}=\frac{1}{\sqrt{2}}(\ket{H}_A\ket{V}_B+\ket{V}_A\ket{H}_B),  
\end{align}
is the polarization maximally entangled state. 
From the coefficient in Eq.~(\ref{eq:swapping}), we find that the ideal success probability of our protocol is $|\alpha|^2|\gamma|^2\sqrt{\eta_{\rm C}}$, i.e. it scales with $\sqrt{\eta_{\rm C}}$ for channel transmission while the rate of directly transmitting polarization entangled photons from Alice to Bob is propotional to $\eta_{\rm C}$. 



The factor $|\alpha|^2|\gamma|^2$ in the above success probability reflects the fact that the generated polarized photon pair consists of one photon from the TMSV and another photon from the coherent state. In practice, however, the higher order photons of the TMSV and the coherent state may also be included and contribute to degrade the fidelity to the ideal entanglement, where the leading terms of occuring these unwanted events are $|\gamma|^4\sqrt{\eta_{\rm C}}$ and $|\alpha|^4|\gamma|^2\sqrt{\eta_{\rm C}}$, respectively. 
To avoid them, therefore, $1 \gg |\alpha|^2 \gg |\gamma|^2$ must be satisfied. 
See Ref.~\cite{supplemental} for more details of the protocols and the effect of the multiphotons from the TMSV and the coherent state.

{\it Experimental setup.$-$}
\begin{figure*}[t]
    \centering
    \includegraphics[width=1\linewidth]{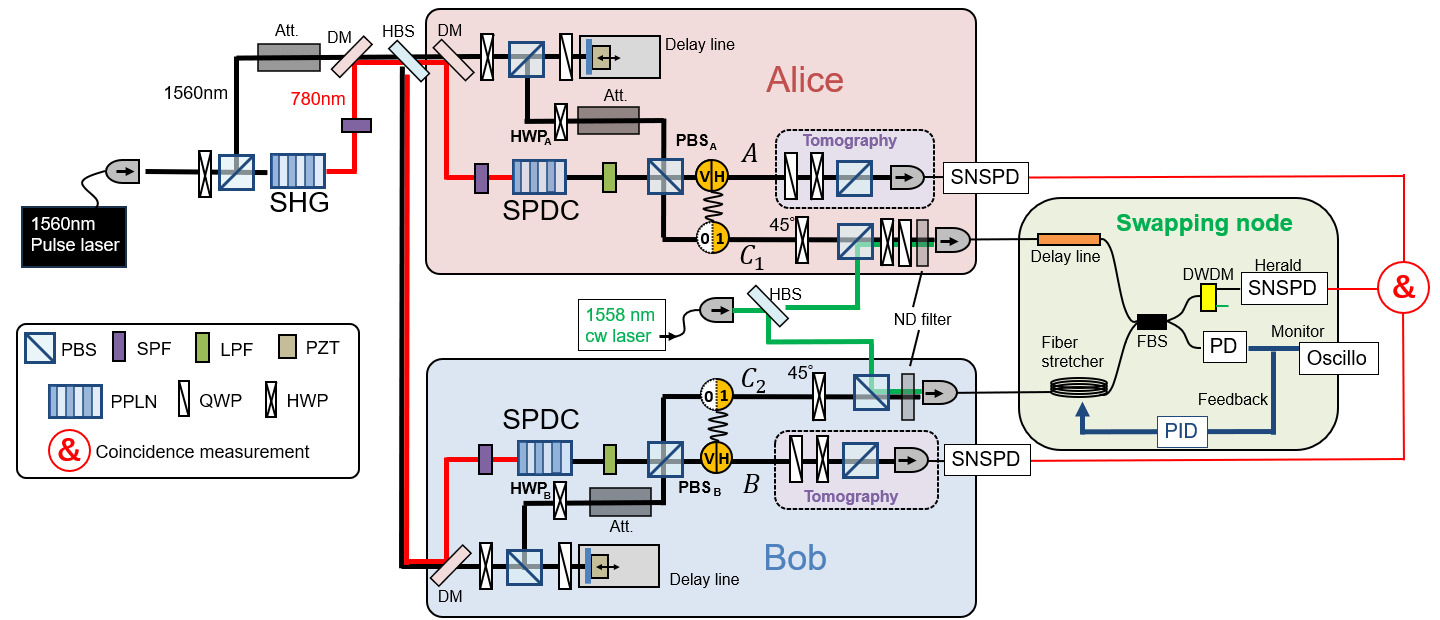}
    \caption{\raggedright Experimental setup. PBS: polarizing beam splitter, SPF: short-pass filter, DM: dichroic mirror, HBS: half beam splitter, QWP: quarter-wave plate, HWP: half-wave plate, PPLN: periodically poled lithium niobate, LPF: long-pass filter, ND filter: neutral density filter, FBS: fiber beam splitter, DWDM: dense wavelength division multiplexer, and PD: photodiode. }
    \label{fig3}
\end{figure*}
Our experimental setup is shown in Fig.~\ref{fig3}.
For generating hybrid entangled states at Alice and Bob, 
we use a mode-locked fiber laser with the central wavelength of \SI{1560}{nm}, 
the repetition rate of \SI{1}{GHz} and the pulse width of \SI{4.9}{ps}.
After a portion of the laser light is used for the second harmonic generation~(SHG) by a Type-0 periodically poled lithium niobate~(PPLN) waveguide, the resulting SH light at \SI{780}{nm} and the attenuated fundamental laser light at \SI{1560}{nm} are distributed to Alice and Bob. 

At Alice and Bob, the lights are separated by a dichroic mirror~(DM). 
The SH light is used as a pump light to generate the TMSV in Eq.~(\ref{eq:TMSV}) 
at a Type-II PPLN waveguide. 
The V-polarized weak coherent light at \SI{1560}{nm} written in Eq.~(\ref{eq:wcp}) is mixed with the SPDC photons at a PBS after passing through a delay line and an optical attenuator~(Att). 
This results in the hybrid entangled states $\ket{\psi}_{AC_1}$ at Alice's side, 
and $\ket{\psi}_{BC_2}$ at Bob's side. 
At each side, the mean photon number of the weak coherent light and excitation probability of SPDC are set to  $|\alpha|^2\sim 0.10$ and $|\gamma|^2\sim 6.0 \times 10^{-3}$, respectively.

The photons at modes $C_1$ and $C_2$ are attenuated by variable neutral density (ND) filters, that act as the channel losses, and then sent to a swapping node through single-mode fibers. 
The photons are mixed at a fiber-based BS~(FBS), and then they are measured by a photon detector connected to one of the output ports of the FBS. Heralded by this photon detection, the polarization-entangled state in modes A and B is shared between Alice and Bob. 
To maximize the fidelity of the entanglement swapping, an optical delay line is inserted into the fiber for synchronizing the arrival time of the wave packets. 
In addition, a fiber stretcher is installed for phase stabilization, which is stabilized by 
the reference light centered at \SI{1558}{nm}, 
traveling along two separate paths originating from Alice and Bob.
The reference light is removed by a DWDM at one of the output ports of the FBS where the photon detection is performed. 
At the other output port, the reference light is detected by a photodiode, and the resulting signal is used for PID feedback to stabilize the optical path length.

Each of the two photons at modes A and B is measured by a photon detector equipped with a quarter-wave plate~(QWP), a half-wave plate~(HWP), and a PBS for quantum state tomography~\cite{James01}. In all experiments, we use superconducting nanowire single-photon detectors~(SNSPDs) with their quantum efficiencies of $\sim \SI{80}{\%}$. 
The three photons measured by the SNSPDs are spectrally filtered using fiber-based bandpass filters with the bandwidths of \SI{0.4}{nm}. 

\begin{figure}
    \centering
    \includegraphics[width=0.9\linewidth]{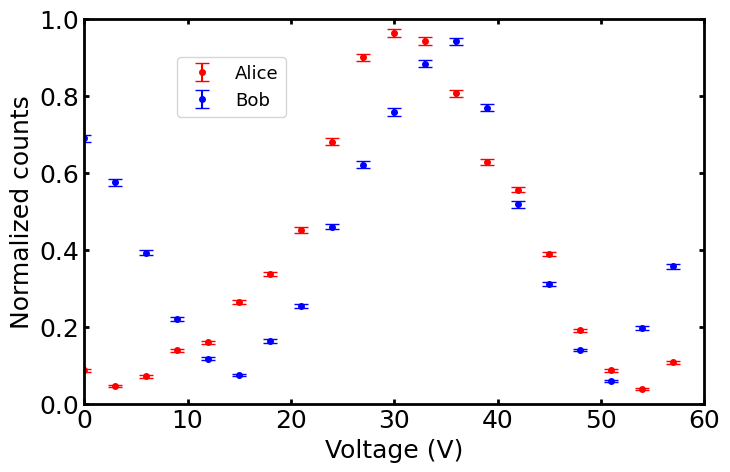}
    \caption{\raggedright The observed coincidence counts between $A~(B)$ and $C_1$~($C_2$) at Alice~(Bob), plotted in red~(blue). 
    The horizontal axis represents the voltage applied to the PZT mounted on the free-space delay line, which corresponds to phase $\theta$. 
    }
    \label{fig4}
\end{figure}

{\it Evaluation of the hybrid entanglement.$-$}
The quality of the hybrid entanglement source is mainly determined by the mode overlap between the SPDC photons and the weak coherent light. 
For Alice's source, this is evaluated by turning off Bob's pump and rotating HWP$_A$ such that the portion of the coherent light, $\ket{0_V}_{A}+\alpha\ket{1_V}_{A}$, is in the H-polarization, $\ket{0_H}_{C_1}+\beta\ket{1_H}_{C_1}$. 
The SPDC photons and the weak coherent light are mixed at PBS$_A$. 
Then conditioned on the photon detection at mode $C_1$, the state in mode $A$ before the HWP and the PBS (in the box of ``Tomography'') is $\ket{\phi}_A=|\alpha||\beta|\ket{V}_A+e^{i\theta}|\gamma|\ket{H}_A$ if they are perfectly overlapped, where $\theta=\varphi_\gamma-\varphi_\alpha-\varphi_\beta$ ($\varphi_\gamma, \varphi_\alpha, \varphi_\beta$ are the phase of $\gamma$, $\alpha$ and $\beta$). 
Then the visibility of the first and second terms of $\ket{\phi}_A$ is observed by projecting them onto the diagonal polarization by the HWP and the PBS. 
In this setting, with $|\alpha||\beta|=|\gamma|$, the coincidence probability of modes $A$ and $C_1$ directly reflects the overlap of the SPDC photons and the weak coherent light. 
This also works for Bob's source. 
Figure~\ref{fig4} shows the experimental results of the oscillation of the coincidence probabilities between $A$ and $C_1$ at Alice and 
$B$ and $C_2$ at Bob. 
Note that the coincidence counts from the multi-photon terms of the SPDC and coherent states are pre-measured and removed from the figure. 
Then the visibilities, i.e., the mode overlaps, are estimated as $M_A=0.924\pm 0.004$ and $M_B=0.881\pm 0.006$, respectively.
See Ref.~\cite{supplemental} for the details of this evaluation.

\if
{\it Evaluation of mode overlap between SPDC photons and coherent light.$-$}
To evaluate the quality of the hybrid entangled states initially prepared at Alice and Bob, we measured the mode overlap between the SPDC photon and the weak coherent light at each location. 
At Alice's side, we rotated $\rm HWP_A$ slightly and prepared H-polarized weak coherent light written by $\ket{\beta}_{C_1}=\ket{0_H}_{C_1}+\beta\ket{1_H}_{C_1}$. 
Using the HWP and the PBS in $C_1$, the H-polarized light and 
the V-polarized number-state photon forming $\ket{\psi}_{AC_1}$ in Eq.~(\ref{initial}) are mixed as the diagonally polarized photons. 
When there is no mode mismatch of the photons in mode $C_1$, the photon detection after the PBS generates an unnormalized state in mode $A$ as 
\begin{equation}
    \label{telep}
    \ket{\phi}_A=|\alpha||\beta|\ket{V}_A+e^{i\theta}|\gamma|\ket{H}_A,
\end{equation}
where $\theta=\varphi_\gamma-\varphi_\alpha-\varphi_\beta$ ($\varphi_\gamma, \varphi_\alpha, \varphi_\beta$ are the phase of $\gamma$, $\alpha$ and $\beta$).
By satisfying the condition $|\alpha||\beta|=|\gamma|$, 
a probability of projecting photon $A$ onto the diagonal polarization ideally exhibits full oscillation from 0 to 1 depending on the phase, under perfect mode matching between the H-polarized SPDC photon and the V-polarized coherent light in mode $A$. 
In practice, however, mode mismatch between the SPDC photons and the coherent light in both modes $A$ and $C_1$ reduces the visibility of the oscillation. 
Assuming that mode overlap probabilities in modes $A$ and $C_1$ are identical and denoted by $M$, the coincidence probability between these modes is given by 
$P_\theta \propto (1+M\cos\theta)/2$. 
Therefore, the oscillation visibility determined by $(P_0-P_\pi)/(P_0+P_\pi)$
exactly corresponds to $M$. 
This is also the case on Bob’s side.

The experimental results of the oscillation of the coincidence probabilities between $A$ and $C_1$ at Alice and 
$B$ and $C_2$ at Bob are shown in Fig.~\ref{fig4}. 
*** 
Some counts are from multi-photon terms whose coefficients are $|\alpha||\gamma|,|\beta||\gamma|$, and they decrease the visibility.
We pre-measured the counts from these terms and plotted the counts with their effects removed.
From the results, the obtained visibilities, namely the mode overlap, are $M_A=0.924\pm 0.004$ and $M_B=0.881\pm 0.006$, respectively.
\fi

{\it Entanglement swapping.$-$}
First, we fixed the channel transmittance $\eta_{\rm C}$ in modes $C_1$ and $C_2$ to be 0.066 and performed the single-click entanglement swapping. 
For the successful events, we reconstructed the density matrix of the polarization entagled state by the two-qubit quantum state tomography. 
The channel transmittance $\eta_{\rm C}$ includes all losses in modes $C_1$ and $C_2$ except for the detection efficiency $\eta_{\rm D}$ of the detectors. 
$\eta_{\rm D}$ includes the efficiency of the bandpass filter followed by the SNSPD and is around 0.12 for all detectors. 
The channel transmittances for the local channel at Alice's and Bob's sides, are $\eta_{\rm LC}\sim 0.2$. 
\if
We distributed polarization entanglement via single-click entanglement swapping using hybrid entangled states.
We first performed two-qubit quantum state tomography after the entanglement swapping to verify the correct operation of the protocol. 
In the experiment, the channel transmittance $\eta_{\rm CH}$ determined 
by local collection efficiencies (mainly determined by the fiber coupling efficiency and the loss at the fiber adapters) of the number-state photons in $C_1$ and $C_2$
was 0.066. 
\fi
\begin{figure}[b]
    \centering
    \includegraphics[width=1\linewidth]{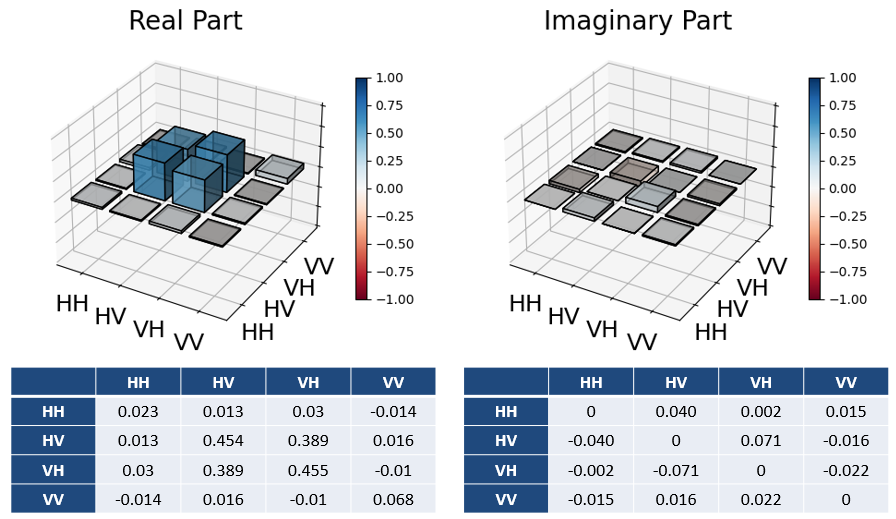}
    \caption{\raggedright Reconstructed density matrix of the distributed state. 
     The measurement time for each polarization setting was \SI{15}{s}.
     The distribution rate was \SI{6.6}{Hz}. 
     }
    \label{fig6}
\end{figure}
The reconstructed density matrix $\rho$ is shown in Fig.~\ref{fig6}.
The fidelity to the ideal polarization entangled state, defined as $\langle\Psi^+_{\rm pol}|\rho|\Psi^+_{\rm pol}\rangle$,
is $0.843\pm0.074$, which shows the succesful entanglement distribution between Alice and Bob. 
From the theoretical model including the mode mismatches of the hybrid entangled states, the fidelity is estimated to be  $(1 + M_A M_B)/2 = 0.907\pm 0.003$, which fits with the experimental result and shows that the main imperfection comes from the mode mismatch in the hybrid sources. 

\begin{figure}[t]
    \centering
    \includegraphics[width=1\linewidth]{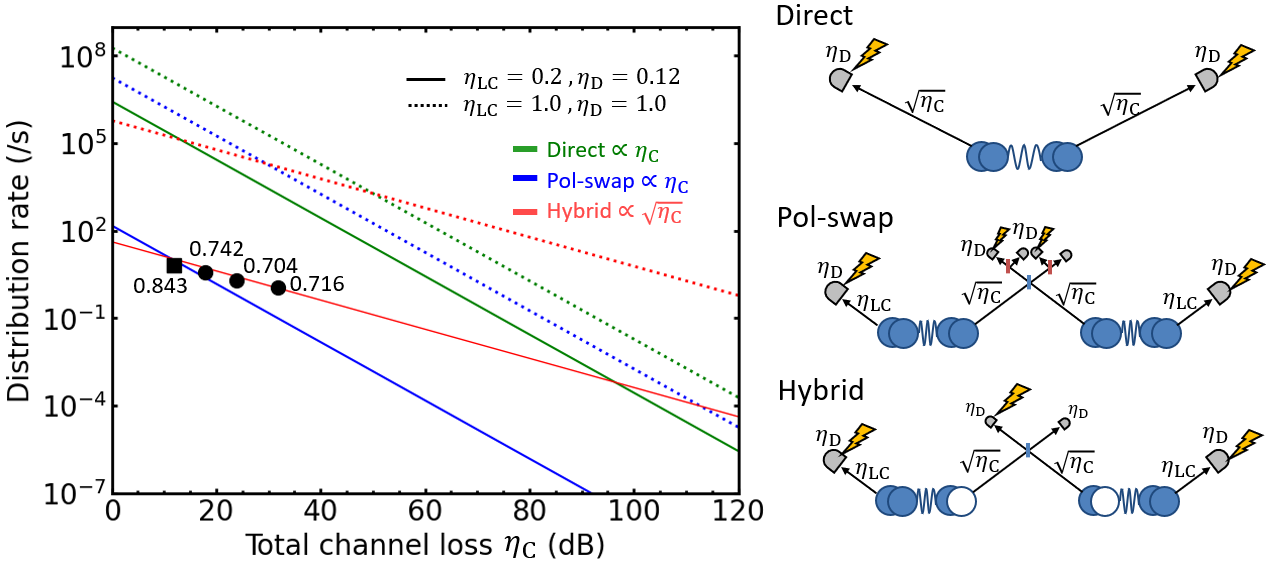}
    \caption{\raggedright Rate-loss scaling. 
    Green, blue, and red lines show direct transmission protocol, protocol based on two-photon interference, and our protocol (hybrid), respectively. Black plots are experimental data. The numbers near the plots indicate the fidelity for the square plot and the lower bound of the fidelity for the circle plots.
    Including error bars, the values are $0.742\pm0.048$, $0.704\pm0.070$, and $0.716\pm0.058$, respectively, from left to right.
    The solid lines use 
    $\eta_{\rm LC}=0.2$ and $\eta_{\rm D}=0.12$,
    and dotted ones use 
    $\eta_{\rm LC}=\eta_{\rm D}=1.0$. 
    }
    \label{fig7}
\end{figure}

\if
Assuming that mode mismatches in the hybrid entangled states described in the previous paragraph are the sole imperfection, the theoretical fidelity is calculated as $(1 + M_A M_B)/2 = 0.907\pm 0.003$, which is in good agreement with the observed value within the margin of error.
is slightly higher than the experimentally observed value.
*** We consider the causes of the difference from the experimental value to be the mode and polarization mismatch of flying qubits, phase drift, loss in channel, and multi-photon terms of hybrid entanglements.
\fi
Next, we performed a similar experiment under various channel losses in paths $C_1$ and $C_2$. 
Figure~\ref{fig7} plots the distribution rates (coincidence count rates) as a function of the channel loss. 
The number at the squared plot is the fidelity whereas the numbers at the circled plots are the lower bound of the fidelity. 
The lower bound of the fidelity~\cite{Nagata02,Ikuta11} is estimated as $F_{\rm LB}=(-V_{ZZ}+V_{XX})/2$, instead of performing full tomography, where the visibilities $V_{ZZ}=\expval{Z_AZ_B}$ and $V_{XX}=\expval{X_AX_B}$ are determined based on the Pauli operators $Z=\ketbra{H}{H}-\ketbra{V}{V}$ and $X=\ketbra{+}{+}-\ketbra{-}{-}$, where $\ket{\pm}=(\ket{H}\pm\ket{V})/\sqrt{2}$. 
The plots clearly show that the rate-loss scaling proportional to $\sqrt{\eta_{\rm C}}$. 

\if
To shorten the measurement time for the high-loss experiments, 
we estimated a lower bound on the fidelity~\cite{Nagata02,Ikuta11}, $F_{\rm LB}=(-V_{ZZ}+V_{XX})/2$, instead of performing full tomography, where the visibilities $V_{ZZ}=\expval{Z_AZ_B}$ and $V_{XX}=\expval{X_AX_B}$ are determined based on the Pauli operators $Z=\ketbra{H}{H}-\ketbra{V}{V}$ and $X=\ketbra{+}{+}-\ketbra{-}{-}$, where $\ket{\pm}=(\ket{H}\pm\ket{V})/\sqrt{2}$. 
Fig.~\ref{fig7} shows the dependence of the observed count rates and fidelities on the channel loss. 
It is clearly observed that the distribution rate of the high-fidelity polarization-entangled state scales proportionally to $\sqrt{\eta}$. 
\fi

The distribution rate of our protocol $R_{\rm hybrid}$ is estimated to be 
\begin{equation}
\label{rate}
    R_{\rm hybrid} = |\alpha|^2|\gamma|^2\sqrt{\eta_{\rm C}}\times\eta_{\rm LC}^2\times\eta_{\rm D}^3\times f_{\rm rep},
\end{equation}
where $f_{\rm rep}=\SI{1}{GHz}$ is the repetition rate and the other parameters in the experiment are $|\alpha|^2\sim 0.10$, $|\gamma|^2\sim 6.0\times10^{-3}$, $\eta_{\rm LC}\sim 0.20$, and $\eta_{\rm D}\sim 0.12$. 
This is indicated by the red solid line in Fig.~\ref{fig7}, which shows good agreement with the experimental data. 
As a reference, the ideal performance of this protocol without unwanted optical losses in the system, calculated using 
$\eta_{\rm LC}=\eta_{\rm D}=1.0$, is shown as a red dotted line. 

In Fig.~\ref{fig7}, the result is compared with the standard entanglement swapping using two-photon interference~\cite{Hong87} between polarization-entangled photon pairs~\cite{Jin15, Tsujimoto18a, Zhan23, Zhan25}.   
Its rate is given by $R_{\rm pol-swap} = \frac{1}{2}R_{\rm gen}^2 \eta_{\rm C}\eta_{\rm D}^4\eta_{\rm LC}^2f_{\rm rep}$, where $R_{\rm gen}$ is the photon-pair generation probability for the SPDC which directly corresponds to $|\alpha|^2$ and $f_{\rm rep}$ is the repetition rate of the pump pulse. 
For fare comparison, these parameters are chosen to be the same as that of our experiment. 
This is indicaed by the blue solid line in the figure. 
Our result is outperforming the standard entanglement swapping as well as showing the square-root advantage of the rate-loss scaling. 
We also made a comparison with the direct transmission of the polarization-entangled photon pairs. 
Its rate is given by $R_{\rm direct} = R_{\rm gen} \eta_{\rm C}\eta_{\rm D}^2 f_{\rm rep}$, which is indicated by the green solid line. 
Although our result is still below this line, they cross at the high-loss regime ($\sim 100~{\rm dB}$). 
Moreover, in principle, the performance of our protocol can be largely increased by improving the local channel transmittance and the detection efficiency. The rates with $\eta_{\rm LC}=\eta_{\rm D}=1$ are compared by the dotted lines, where we observed that the crossover of our protocol and the other protocols occur at significantly lower loss regime. 

\if
We compare the rates of the protocols for distributing polarization entanglement with and without hybrid entanglement, 
under the condition that the excitation probabilities are chosen so that the effect of multiple-photon emissions on the fidelity of the distributed state is roughly the same. The comparison is shown in Fig.~\ref{fig7}. 
We calculated the rate of the direct transmission of a polarization-entangled photon pair without entanglement swapping as 
$R_{\rm direct} = R_{\rm gen} \eta_{\rm CH}\eta_{\rm DE}^2 f_{\rm rep}$. 
The excitation probability of the SPDC-based polarization-entangled photon pair, denoted by $R_{\rm gen}$, was set to $|\alpha|^2\sim 0.10$ in this analysis. 
We consider the degradation of fidelities is from the increased effect of phase drift due to the longer measurement time and that we did not update the parameters $\alpha,\gamma$ for the loss.
The optimal probability of transmitting photons depends on the channel loss.
Another protocol for comparison is based on the entanglement swapping using two-photon interference~\cite{Hong87} between polarization-entangled photon pairs~\cite{Jin15, Tsujimoto18a, Zhan23, Zhan25}. 
The rate was calculated by $R_{\rm 2} = \frac{1}{2}R_{\rm gen}^2 \eta_{\rm CH}\eta_{\rm DE}^4\eta_{\rm LCE}^2f_{\rm rep}$. 
From the comparison, our protocol based on the hybrid entangled photon pairs shows better scaling and outperforms in the high-loss regime. 
\fi

{\it Bell test.$-$}
Finally, we performed the CHSH-type Bell test~\cite{Clauser69} on the distributed polarization-entangled photons as a demonstration of quantum network application. 
In this experiment, we removed ND filters in front of the fiber collimators.
Alice's measurement bases are $Q:\{\ket{H}, \ket{V}\}$ and $R:\{\ket{+}, \ket{-}\}$, and Bob's are $S:\{\ket{\theta},\ket{\theta+90^\circ}\}$ and $T:\{\ket{-\theta}, \ket{-\theta+90^\circ}\}$, where $\ket{\theta} = \cos{\theta}\ket{H}+\sin{\theta}\ket{V}$. 
For each measurement basis, we assign $+1$ to the first basis vector and $-1$ to the second.
The CHSH parameter $S$ is determined by
\begin{equation}
    \label{CHSHscore}
    S = |\langle QS \rangle + \langle QT \rangle + \langle RS \rangle - \langle RT \rangle|.
\end{equation}
From the simulation of the Bell test on the reconstructed density matrix $\rho$ in Fig.~\ref{fig6}, the $S$ parameter is expected to reach its maximum value of $2.313$ at $\theta=-21.5^\circ$. 
Based on this simulation result, we carried out the Bell test in an actual experiment at the angle. 
The observed value of the $S$ parameter was $2.302 \pm 0.066$, which is in good agreement with the simulation result. 
The experimental value clearly violates the upper bound of $2$ predicted by any local hidden variable theory, by approximately 5 standard deviations. 

{\it Discussion.$-$ }
Our result can be extended to the multipartite entanglement distribution scenario. 
In Refs.~\cite{Roga23, Shimizu25}, protocols for efficiently distributing photon-number based multipartite entangled states, such as W-, Dicke-, and GHZ-states, are proposed. 
Combining these with our hybrid approach enables the distribution of polarization-based multipartite entangled states with better rate-loss scaling. The advantage compared to the direct transmission is more prominent than that of the bipartite case~\cite{supplemental}. 
Especially, we find that an application of our hybrid approach to the efficient GHZ-state distribution scenario can remedy the problem of multi-photon effects observed in Ref.~\cite{Shimizu25} by the entanglement distillation-like effect. See Ref.~\cite{supplemental} for its details.

In conclusion, we demonstrate a protocol of efficiently distributing polarization entanglement by using the hybrid entanglement sources and the single-click entanglement swapping. 
The distributed state shows high fidelity to the ideal polarization Bell state and we experimentally observe the square-root improvement of the rate-loss scaling from the conventional protocols. 
In addition, our technique is directly applicable to the protocol of efficiently distributing multipartite polarization entangled states. 
We believe that this study will accelerate research into large-scale quantum network applications.

\begin{acknowledgements}
{\it Acknowledgements.$-$ }R.I. and M.T. acknowledge the members of the Quantum Internet Task Force for the comprehensive and interdisciplinary discussions on the quantum internet.
This work was supported by JST CREST, JPMJCR24A5; MEXT Q-LEAP, JPMXS0118067395; Center for Spintronics Research Network (CSRN), Keio University; NEXT Leading Initiative for Excellent Young Researchers; Program for the Advancement of Next Generation Research Projects, Keio University; JST CRONOS, JPMJCS24N6; JST ASPIRE, JPMJAP2427; JST Moonshot R\&D, JPMJMS2061, JPMJMS226C, JPMJMS2066; R \& D of ICT Priority Technology Project JPMI00316; and FOREST Program, JST JPMJFR222V.
\end{acknowledgements}

\clearpage
\widetext
\begin{center}
   \textbf{\large Supplemental Material: Improving the Rate-Loss Scaling in Polarization Entanglement Distribution using Single-Click Entanglement Swapping} 
\end{center}
\setcounter{equation}{0}
\setcounter{figure}{0}
\setcounter{table}{0}
\setcounter{page}{1}

\section{Explicit description of the multiphoton effects}
Here, we examine the impact of multiphoton components on the protocol. 
As described in the main text, the hybrid entangled state is generated by mixing a two-mode polarization squeezed vacuum~(TMSV) state with a vertically polarized weak coherent state via a polarizing beam splitter~(PBS). 
Since both the TMSV and the coherent state inherently contain multiphoton components, noise due to these components is fundamentally unavoidable.

The TMSV state from the type-$\mathrm{I}\hspace{-1.2pt}\mathrm{I}$ spontaneous parametric down conversion~(SPDC) sources is given by,
\begin{equation}
    \label{tmsv}
    \ket{\rm TMSV} = \sqrt{1-\gamma^2} \sum_{n=0}^\infty \gamma^n\ket{n_Hn_V},
\end{equation}
where $\gamma$ is the squeezing parameter. 
The weak coherent state that each user prepares in vertically polarized mode is,
\begin{equation}
    \label{wcs}
    \ket{\alpha}=e^{-\frac{|\alpha|^2}{2}}\sum_{n=0}^\infty \frac{\alpha^n}{\sqrt{n!}}\ket{n_V}.
\end{equation}
When these two states are input into a PBS from different ports (see Fig.~\ref{fig0}(a)), the output is,
\begin{equation}
    \label{hybrid}
    \begin{split}
        &\ket{\phi}_{AC_1} =\ket{\alpha_V}_{A}\otimes \ket{\rm TMSV}_{A{C_1}}\\
        =&e^{-\frac{|\alpha|^2}{2}}\sqrt{1-\gamma^2}[\ket{0_V0_H}_A\ket{0_V}_{C_1} + \alpha\ket{1_V0_H}_A\ket{0_V}_{C_1} + \gamma\ket{0_V1_H}_A\ket{1_V}_{C_1}\\ + &\frac{\alpha^2}{\sqrt{2}}\ket{2_V0_H}_A\ket{0_V}_{C_1} + \gamma^2\ket{0_V2_H}_A\ket{2_V}_{C_1} + \alpha\gamma\ket{1_V1_H}_A\ket{1_V}_{C_1}+\cdots].
    \end{split}
\end{equation}
\renewcommand{\thefigure}{S\arabic{figure}} 
\setcounter{figure}{0}
\begin{figure}[h]
    \centering
    \includegraphics[width=0.8\linewidth]{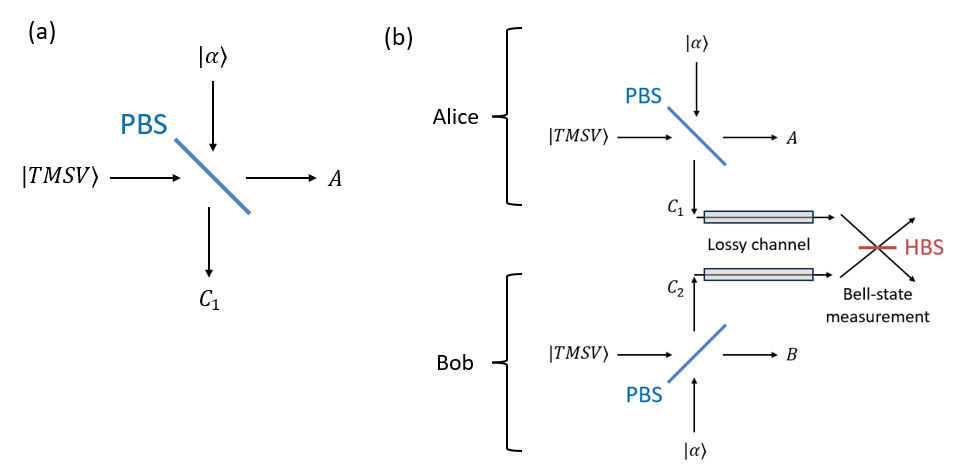}
    \caption{Schematic of the experiment. (a)Hybrid entanglement generation. (b)Entanglement swapping.}
    \label{fig0}
\end{figure}
Hereafter, we omit the normalization term $e^{-\frac{|\alpha|^2}{2}}\sqrt{1-\gamma^2}$, as it is close to 1. 
This is the initial state generated locally, where the second and third terms correspond to the unnormalized hybrid entanglement,
\begin{equation}
\label{hybrid_ent}
    \alpha\ket{V0}_{AC_1}+\gamma\ket{H1}_{AC_1}.
\end{equation}
Bob also generates the same state $\ket{\phi}_{BC_2}$.
Each user transmits their photon-number qubit~(mode $C_1$ and $C_2$) to the swapping node via lossy channels.
At the swapping node, modes $C_1C_2$ are projected to $\ket{\Psi^+_{01}}_{C_1C_2}=\frac{1}{\sqrt{2}}(\ket{10}_{C_1C_2}+\ket{01}_{C_1C_2})$.
Due to the loss during transmission, the initial state $\ket{\phi}_{AC_1}\otimes\ket{\phi}_{BC_2}$ becomes mixed. 
Since only specific components contribute to the projection measurement, we focus exclusively on them in what follows.
Since only vertically polarized photons exist, we omit the polarization subscripts of mode $C_1$ and $C_2$ hereafter.
Successful projection onto $\ket{\Psi^+_{01}}_{C_1C_2}$ is heralded by the detection of a single photon at either output port after interfering modes $C_1$ and $C_2$ on a fiber beam splitter.
Among the terms in $\ket{\phi}_{AC_1}\otimes\ket{\phi}_{BC_2}$ that can lead to this event, the desired one are given by $\alpha\gamma\ket{1_V0_H}_{A}\ket{0_V1_H}_{B}\ket{01}_{C_1C_2}$ and $\alpha\gamma\ket{0_V1_H}_{A}\ket{1_V0_H}_{B}\ket{10}_{C_1C_2}$, with the success probability of the single-photon transmission $\sqrt{\eta_{\rm C}}$.
Thus, the heralding probability from ideal terms is,
\begin{equation}
    P_{\rm ideal}= \sqrt{\eta_{\rm C}}|\alpha|^2|\gamma|^2.
\end{equation}
On the other hand, while multiple undesired terms exist, the dominant contribution is expected from,
\begin{align}
\begin{split}
    &\frac{\alpha^2\gamma}{\sqrt{2}}(\ket{2_V0_H}_{A}\ket{0_V1_H}_{B}\ket{01}_{C_1C_2}+
    \ket{0_V1_H}_{A}\ket{2_V0_H}_{B}\ket{10}_{C_1C_2}),\\
    &\alpha^2\gamma(\ket{1_V1_H}_{A}\ket{1_V0_H}_{B}\ket{10}_{C_1C_2}+ \ket{1_V0_H}_{A}\ket{1_V1_H}_{B}\ket{01}_{C_1C_2}),\\
    &\gamma^2\ket{0_V1_H}_{A}\ket{0_V1_H}_{B}\ket{11}_{C_1C_2}.\\
\end{split}    
\end{align}
These terms contribute to the heralding with the following probability,
\begin{equation}
\begin{split}
    P_{\rm noise}&=\frac{3}{2}\sqrt{\eta_{\rm C}}|\alpha|^4|\gamma|^2+[1-(1-\sqrt{\eta_{\rm C}})^2]|\gamma|^4\\
    &=\sqrt{\eta_{\rm C}}|\gamma|^2[\frac{3}{2}|\alpha|^4+(2-\sqrt{\eta_{\rm C}})|\gamma|^2].
\end{split}
\end{equation}
Assuming that we do not have any other imperfections, the fidelity to the ideal state $|\Psi^+_{\rm pol}\rangle=\frac{1}{\sqrt{2}}(\ket{HV}+\ket{VH})$ is,
\begin{equation}
\label{fidelity}
    F = \frac{P_{\rm ideal}}{P_{\rm ideal}+P_{\rm noise}} = \frac{1}{1+\frac{3}{2}|\alpha|^2+\frac{(2-\sqrt{\eta_{\rm C}})|\gamma|^2}{|\alpha|^2}}.
\end{equation}
Equation~(\ref{fidelity}) indicates that the following condition is necessary to achieve high fidelity.
\begin{equation}
    |\gamma|^2 \ll |\alpha|^2 \ll 1.
\end{equation}
Here, the fidelity is defined as $\langle\Psi^+_{\rm pol}|\rho_{AB}|\Psi^+_{\rm pol}\rangle$, and the distributed state $\rho$ can be described as,
\begin{equation}
    \rho_{AB} = P_{\rm ideal}|\Psi^+_{\rm pol}\rangle\langle\Psi^+_{\rm pol}| + \sqrt{\eta_{\rm C}}|\alpha|^4|\gamma|^2\ketbra{\phi_1} + 2\sqrt{\eta_{\rm C}}|\alpha|^4|\gamma|^2\ketbra{\phi_2} + \sqrt{\eta_{\rm C}}(2-\sqrt{\eta_{\rm C}})|\gamma|^4\ketbra{\phi_3},
\end{equation}
where, 
\begin{align}
    \begin{split}
        \ket{\phi_1}_{AB}&=\frac{1}{\sqrt{2}}(\ket{1_H}_A\ket{2_V}_B+\ket{2_V}_A\ket{1_H}_B),\\
        \ket{\phi_2}_{AB}&=\frac{1}{\sqrt{2}}(\ket{1_H1_V}_A\ket{1_V}_B+\ket{1_V}_A\ket{1_H1_V}_B),\\
        \ket{\phi_3}_{AB}&=\ket{1_H}_A\ket{1_H}_B.\\
    \end{split}
\end{align}

\section{Evaluation of the hybrid entanglement}
To evaluate the quality of the hybrid entangled states initially prepared at Alice and Bob, we measured the mode overlap between the SPDC photon and the weak coherent light at each location. 
At Alice's side, we rotated $\rm HWP_A$ and prepared H-polarized weak coherent light written by $\ket{\beta}_{C_1}=\ket{0_H}_{C_1}+\beta\ket{1_H}_{C_1}$. 
Then, the whole state of Alice is,
\begin{equation}
\label{vis}
\begin{split}
    \ket{\varphi} =&\ket{\rm TMSV}_{A{C_1}}\otimes\ket{\alpha_V}_{A}\otimes\ket{\beta_H}_{{C_1}}\\
    =& e^{-(\frac{|\alpha|^2}{2}+\frac{|\beta|^2}{2})}\sqrt{1-\gamma^2}(\alpha\ket{0_H1_V}_A\ket{0_H0_V}_{C_1}+\beta\ket{0_H0_V}_A\ket{1_H0_V}_{C_1}+\gamma\ket{1_H0_V}_A\ket{0_H1_V}_{C_1}+\\
    &\alpha\beta\ket{0_H1_V}_A\ket{1_H0_V}_{C_1}+\gamma\alpha\ket{1_H1_V}_A\ket{0_H1_V}_{C_1}+\gamma\beta\ket{1_H0_V}_A\ket{1_H1_V}_{C_1}+\cdots).
\end{split}
\end{equation}
Using the HWP and the PBS in $A$ and $C_1$, we projected both modes to the diagonally polarized states and measured the coincidence counts between $A$ and $C_1$.
The coincidence probability can be calculated from,
\begin{equation}
\label{pcc}
    P_{\rm c.c.}=||_{AC_1}\braket{1_D1_D}{\varphi}_{AC_1}||^2=\left|\frac{1}{2}\gamma+\frac{1}{2}\alpha\beta+ {\rm others}\right|^2.
\end{equation}
We ignore the normalization factor as it does not affect the quantity of interest~(the visibility).
Assuming that $\alpha,\beta,$ and $\gamma$ are sufficiently small that the contribution from ``others''in Eq.~(\ref{pcc}) is negligible, $P_{\rm c.c.}$ depends on the phase $\theta$ as follows,
\begin{equation}
\label{ptheta}
    P_{\rm c.c.} = \frac{1}{4}(|\gamma|^2+|\alpha|^2|\beta|^2+2|\gamma||\alpha||\beta|\cos{\theta})\equiv P_{\theta},
\end{equation}
where $\theta=\varphi_\gamma-\varphi_\alpha-\varphi_\beta$ ($\varphi_\gamma, \varphi_\alpha, \varphi_\beta$ are the phase of $\gamma$, $\alpha$ and $\beta$).
By satisfying the condition $|\alpha||\beta|=|\gamma|$, 
the coincidence probability exhibits full oscillation from 0 to 1 depending on the phase, under perfect mode matching between the SPDC photons and the weak coherent lights.  
Assuming that mode overlap probabilities in modes $A$ and $C_1$ are identical and denoted by $M$, the coincidence probability between these modes is given by 
$P_\theta \propto (1+M\cos\theta)/2$. 
Therefore, the oscillation visibility determined by $(P_0-P_\pi)/(P_0+P_\pi)$
exactly corresponds to $M$. 
This is also the case on Bob’s side.

In practice, it is difficult to completely eliminate the coincidence contributions from the ``others'' in Eq.~(\ref{pcc}). Therefore, we estimated the dominant contributions, $\gamma\alpha\ket{1_H1_V}_A\ket{0_H1_V}_{C_1}$ and $\gamma\beta\ket{1_H0_V}_A\ket{1_H1_V}_{C_1}$, by performing projective measurements onto 
$\ket{1_V}_A\ket{1_V}_{C_1}$ and $\ket{1_H}_A\ket{1_H}_{C_1}$, respectively, and subtracted them from the experimental values to evaluate the actual mode overlap.

\section{Phase stabilization of the fiber channel}
                      
\begin{figure}[h]
    \centering
    \includegraphics[width=0.5\linewidth]{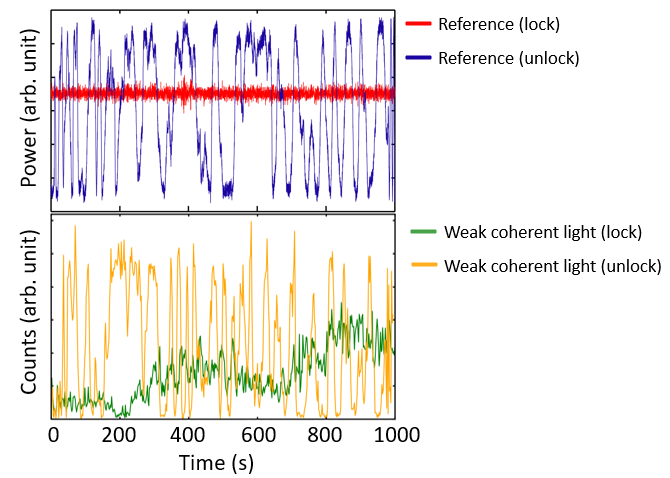}
    \caption{\raggedright Results of the phase stabilization. The upper graph shows the interference of the referential light. The lower graph shows the interference of the weak coherent light at the same wavelength as the photon-number qubit. }
    \label{sfig1}
\end{figure}
We describe the feedback system used to stabilize the relative phase between photons transmitted from Alice and Bob through optical fibers, up to the point where they are combined at the swapping node. 
The reference light for the stabilization was provided by a continuous-wave laser at 1558~nm~(WSL-110, Santec). 
The laser was independent of the laser source~(Ultrafast optical clock, Pritel) used to generate the photon pairs at 1560~nm, and was not synchronized with that laser. 
As shown in Fig.~2 in the main text, the reference light was split and sent to Alice and Bob, and then recombined at the FBS after transmission through optical fibers. The interferometric light from the FBS was detected by a photodiode~(PD). 
The electrical signal from the PD was fed into an analog PID controller~(SIM960, Stanford Research Systems). The feedback signal from the PID controller was then applied to a fiber stretcher~(915B, Evanescent Optics Inc.). 

We show the phase stabilization results in Fig.~\ref{sfig1}.
The upper figure shows the interference of the reference light, and the lower one shows that of the weak coherent light whose wavelength is the same as the flying photons.
Since the reference light is directly controlled, it is well stabilized.
On the other hand, there remains a slow phase drift in the weak coherent light. 
The measurement time is at most 1 minute on each polarization setting. Thus, the drift is not critical in this experiment. Laser synchronization and/or the sample-hold  technique would be helpful for future large-scale experiments.

\section{Distribution of multi-partite entanglement}
\begin{figure}[h]
    \centering
    \includegraphics[width=0.7\linewidth]{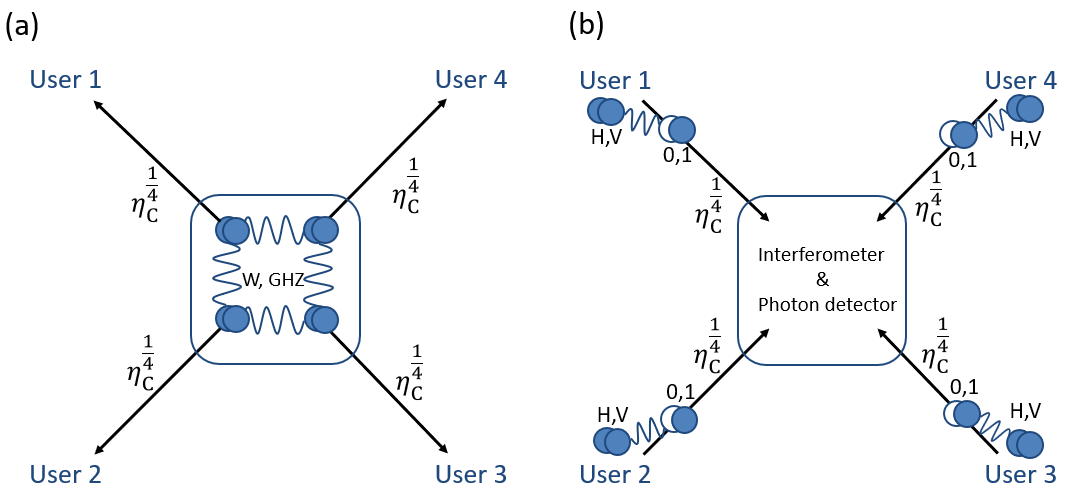}
    \caption{\raggedright Schematic image of multi-partite entanglement distribution. (a)Direct transmission protocol. (b)Protocol extending single-click entanglement swapping. All channel has the same transmittance $\eta_{\rm C}^{\frac{1}{4}}$.} 
    \label{sfig2}
\end{figure}
This study is compatible with a loss-tolerant protocol of multi-partite entanglement distribution proposed in Refs.~\cite{Roga23s, Shimizu25s}.
Here, we consider the distribution of the 4-partite polarization W- and GHZ-states in the star-type quantum network.
All users are connected to the central node through channels whose transmittances are all $\eta_{\rm C}^{1/4}$  (see Fig.~\ref{sfig2}).
A conventional distribution method generates the target state at the central node and sends each photonic qubit to each end-user node.
We refer to this method as ``direct transmission'' because it involves no repeater-like operations.
In this method, all photons must survive the transmission.
Thus, the distribution rate of direct transmission would scale as $\eta_{\rm C}$.

In our protocol (combination of hybrid entanglement and Refs.~\cite {Roga23s, Shimizu25s}), each user prepares a hybrid entanglement Eq.~(\ref{hybrid_ent}) and sends a photon-number qubit to the central node.
At the central node, the incoming photon-number qubits interfere with each other and are detected by single-photon detectors. Depending on the detection pattern, the quantum states remaining in the users' hands are projected accordingly: a single detector click heralds a W-state, while simultaneous clicks in two different detectors herald a GHZ-state.

The distribution rate of this protocol depends on the success probability of getting the target detection pattern and the detection probability of photons in all user nodes.
In the W-state distribution, we need only one photon to be detected at the central node.
One of the four users must detect photons from SPDC, while the other three must detect photons from weak coherent light.
Thus, the distribution rate would be
\begin{equation}
\label{W}
    R_{\rm W}= |\alpha|^6|\gamma|^2\times \eta_{\rm C}^\frac{1}{4} \times f_{\rm rep}.
\end{equation}
Note that we ignore the noramlization factor $e^{-\frac{|\alpha|^2}{2}}\sqrt{1-\gamma^2}$ again because it is close to 1 with the condition $|\gamma|^2\ll|\alpha|^2\ll1$.
Events in which two or more users send photons, but only one is detected due to losses, contribute to noise. To suppress such events, the probability that each user sends a photon — that is, $\gamma/\alpha$ — must be kept small.
In addition, multi-photon in local qubits also contributes to noise, so $\alpha$ must be small as well.
Therefore, as in the main text, $|\gamma|^2 \ll |\alpha|^2 \ll 1$ is also required in the context of multipartite entanglement distribution. 
Moreover, as the number of users increases, the number of noise contributions also increases, requiring the $\gamma/\alpha$ ratio to be further reduced compared to the two-user scenario.
In the GHZ-state distribution, we need two photons to be detected at the central node.
Two of the four users must detect photons from SPDC, while the other two must detect photons from weak coherent light.
Thus, the distribution rate would be
\begin{equation}
\label{GHZ}
    R_{\rm GHZ}\propto  |\alpha|^4|\gamma|^4\times \eta_{\rm C}^\frac{1}{2}\times f_{\rm rep}.
\end{equation}
We use "$\propto$" because the projective measurement of the GHZ-state is probabilistic in Ref.~\cite{Shimizu25s}.
The GHZ-state distribution has the same type of noise as the W-state distribution.
Thus, $|\gamma|^2 \ll |\alpha|^2 \ll 1$ must be kept.

An advantage of the hybrid approach for the GHZ-state distribution is the robustness against the other type of noise, which arises due to the multiple photon detection at the central node. 
The noise comes from the indistinguishability between the two events: the photons come from different users, which is intended, or the photons come from the same user, which is unwanted. 
When the TMSV is used at each user side, these two events occur with the same probability. 
In Ref.~\cite{Shimizu25s}, this issue is solved by replacing the TMSV with the single-photon entangled state, which is prepared by beamsplitting the heralded single-photon conditionally generated from the TMSV. 
In contrast, if we use hybrid entanglement as the local entanglement, the probability of the unwanted event is inherently suppressed compared to the intended one, which allows a drastic simplification of the setup compared to the above single-photon entangled state, and thus is more feasible with the current technology.  
\if
In addition, a distinctive type of noise arises due to the necessity of detecting multiple photons at the central node.
This noise originates from the indistinguishability between an undesirable event, in which two photons are sent from a single user, and a desirable event, in which one photon is sent from each of two users.
In the case where two-mode squeezed vacuum (TMSV) states are used as local entanglement resources, these two types of events occur with equal probability. Reference~\cite{Shimizu25s} addressed this issue by spatially separating the heralded single photon into two modes.
In contrast, when using hybrid entanglement as the local entanglement, the probability of the former event is inherently suppressed compared to the latter, and thus, no additional operations are required to mitigate this effect.
\fi

\begin{figure}[t]
    \centering
    \includegraphics[width=0.5\linewidth]{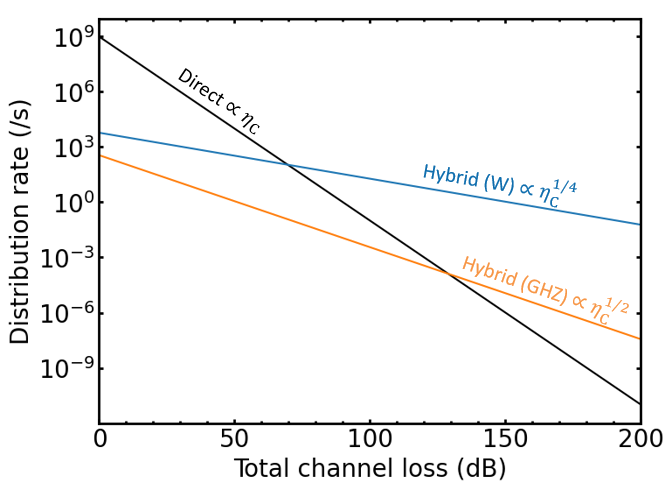}
    \caption{\raggedright Rate-loss scaling of 4-partite entanglement distribution. Direct transmission requires all four photons to survive, and thus, the rate scales as $\eta_{\rm C}$. }
    \label{sfig3}
\end{figure}

Figure~\ref{sfig3} plots $f_{\rm rep}\eta_{\rm C}$, Eq.~(\ref{W}), and Eq.~(\ref{GHZ}) with $|\alpha|^2=0.1$, $|\gamma|^2=6.0\times10^{-3}$, $|\gamma_{\rm SPDC}|^2=0.1$, and $f_{\rm rep}=1.0\times10^9$, that are the same parameters used in the simulation in the main text (Fig.~5). From Fig.~\ref{sfig3}, it is obvious that our protocol has a great advantage in rate-loss scaling. 
Note that no imperfections are included in this figure.

\end{document}